\documentclass[aps,prb,twocolumn]{revtex4-1}
\pdfoutput=1
\usepackage{amsmath}
\usepackage{ amssymb }
\usepackage{graphicx}
\usepackage{color}

\begin{document}

\newcommand{\beq}{\begin{equation}}
\newcommand{\eeq}{\end{equation}}
\newcommand{\beqa}{\begin{eqnarray}}
\newcommand{\eeqa}{\end{eqnarray}}
\newcommand{\note}[1]{{\color{red} [#1]}}
\newcommand{\bra}[1]{\ensuremath{\langle#1|}}
\newcommand{\ket}[1]{\ensuremath{|#1\rangle}}
\newcommand{\bracket}[2]{\ensuremath{\langle#1|#2\rangle}}
\renewcommand{\vec}[1]{\textbf{#1}}
\newcommand{\dagga}{{\phantom{\dagger}}}


\title{Quantum phase transitions  in effective  spin-ladder models\\ for  graphene zigzag nanoribbons}

\author{Cornelie Koop}
\affiliation{Institut f\"ur Theoretische Festk\"orperphysik, JARA-FIT and JARA-HPC, RWTH Aachen University, 52056 Aachen, Germany}

\author{Stefan Wessel}
\affiliation{Institut f\"ur Theoretische Festk\"orperphysik, JARA-FIT and JARA-HPC, RWTH Aachen University, 52056 Aachen, Germany}

\date{\today}

\begin{abstract}
We examine the magnetic correlations in quantum spin models that
were derived recently as  effective low-energy theories for electronic correlation effects on the edge states of  graphene  nanoribbons.
For this purpose, we employ  quantum Monte Carlo
simulations to access the large-distance properties,  accounting for  quantum fluctuations beyond 
mean-field-theory approaches to edge magnetism. 
For certain chiral nanoribbons, 
antiferromagnetic inter-edge couplings
were previously found to induce a gapped quantum disordered  ground state of the effective spin model.  We find that
the extended nature of the  
intra-edge couplings in the effective spin model for zigzag nanoribbons  leads to  a quantum phase transition at a large, finite value of the 
inter-edge coupling.  This quantum critical point separates the quantum disordered region from
a gapless phase of stable edge magnetism at weak intra-edge coupling, which includes the ground states of  spin-ladder models for wide zigzag nanoribbons. 
To study the quantum critical behavior, the effective spin model can be related to a model of two antiferromagnetically coupled Haldane-Shastry spin-half chains with long-ranged ferromagnetic intra-chain couplings. The results for the critical  exponents are  compared also to several recent renormalization group calculations for related long-ranged interacting quantum systems.

\end{abstract}

\maketitle

\section{Introduction}\label{sec:intro}

Graphene-based nanoribbons with zigzag edge termination are characterized by the presence of  an almost flat band of edge states~\cite{Fujita96}. 
The corresponding, strongly increased local density of states allows electron-electron interactions to induce enhanced magnetic correlations along the edges, as compared to bulk graphene~\cite{CastroNeto09}.
In fact, from  a broad range of theoretical studies a general picture has been promoted that the edge states along each edge of the zigzag nanoribbon are gapped out and exhibit a ferromagnetic alignment, thereby forming a pair of edge-superspins which are correlated antiferromagnetically across the nanoribbon's transverse extend~\cite{Wakabayashi98,Son06,Pisani07,Yazyev08,Yazyev10,Jung11,Affleck12,Shi17}. 
Even though recent progress in synthesizing graphene zigzag nanoribbons and  controlling the edge alignment  allows to  identify and better characterize the localized edge states~\cite{Tao11, Magda2014, Makarova15, Ruffieux16}, 
a fully conclusive experimental demonstration of such edge magnetism is still not generally agreed upon.   

The above picture is  aggravated by the fact that even within the most simple theoretical approach to  edge magnetism, based on a  local Hubbard model tight-binding description of  graphene nanoribbons, 
it has been argued that quantum fluctuations, which are neglected in most mean-field-theory based predictions of the edge magnetism,   suppress the ferromagnetic correlations along the nanoribbon edges~\cite{Hikihara03, Feldner11, Golor13a, Golor13, Golor14}. 
For the case of specific chiral nanoribbons, where zigzag-terminated edge segments are separated by armchair-terminated steps, it was also shown within effective quantum spin models for the  magnetic  correlations~\cite{Schmidt13,Koop15}  that the antiferromagnetic inter-edge coupling leads to a quantum disordered state, characterized 
by  an exponential decay of the magnetic correlations along the edges and  a finite spin excitation gap~\cite{Golor14}. 

The effective quantum spin models referred to  above can be derived from the parent Hamiltonian (the Hubbard model on the nanoribbon lattice) via a sequence of controlled approximations that separate on the microscopic level the edge states from the bulk states of the nanoribbon in an optimized Wannier-basis (for details on the derivation of the effective spin model, and the extension to a second-order treatment within the Schrieffer-Wolff transformation, we refer to Refs.~\onlinecite{Schmidt13, Koop15}).
These effective theories are formulated in terms of  a spin-half Heisenberg model 
and the corresponding lattice geometry is that of an effective two-leg ladder  with extended ferromagnetic exchange interactions along the  legs (each representing one of the nanoribbon edges) and  extended antiferromagnetic interactions between spins on different legs. 
While the general form of such  spin-ladder  models for graphene nanoribbons has been described previously~\cite{Wakabayashi98,Yazyev08,Yoshioka03}, the calculations in Refs.~\onlinecite{Schmidt13,Koop15} 
provide a systematic way to evaluate the effective exchange couplings for a given  specific microscopic nanoribbon geometry. 

A useful aspect of such effective  spin-ladder  models is the fact that they allow to probe  long-ranged magnetic correlations on significantly larger length scales than accessible to direct simulations~\cite{Golor13} of the parent Hamiltonian for  chiral nanoribbons in terms of the Hubbard model, so that even large finite correlation lengths can be quantified~\cite{Golor14}. 
For the  chiral ribbons considered in Ref.~\onlinecite{Golor14}, the  interactions in the effective  spin-ladder  model decay exponentially with the spatial distance between the spins. The effective two-leg ladder model therefore  behaves qualitatively similar to  a two-leg ladder Heisenberg model with only  nearest-neighbor ferromagnetic leg coupling and antiferromagnetic rung coupling: any finite value of the rung coupling results in  a gapped quantum disordered state from the  formation of dominant rung singlets~\cite{Roji96,Kolezhuk96,Vekua03}. For the effective spin models with extended interactions, the singlets of the spin gapped state extend over larger spatial regions, quantified by the correlation length~\cite{Golor14}.

\begin{figure}[t]
\includegraphics[width=\columnwidth]{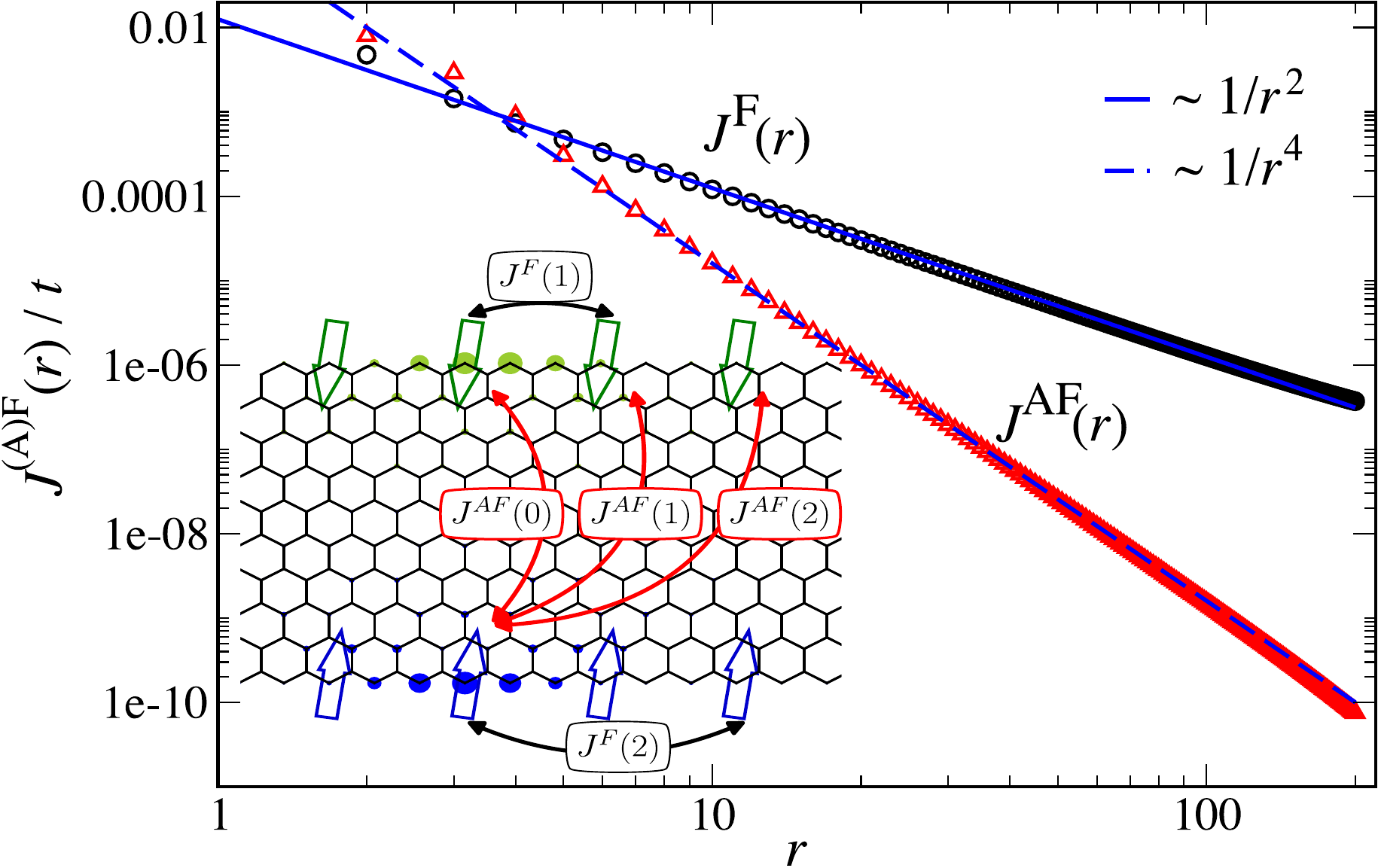}
\caption{(Color online). 
Effective ferromagnetic ($J^\mathrm{F}(r)$) and antiferromagnetic $(J^\mathrm{AF}(r)$)  couplings for a  $W=10$ zigzag nanoribbon for the Hubbard model parameters $U=t$ as a function of  the lateral distance $r$. 
In order to calculate these effective coupling parameters, 
a finite size ribbon with $W=10$ zigzag lines and a total of  $48000$ lattice sites was considered. 
The resulting values for the couplings of  distances $r\leq 4$ are given explicitly in Tab.~\ref{tab:c}.
The inset illustrates a $W=10$ zigzag nanoribbon along with some of  the effective exchange interactions  of the effective  spin-ladder  models in terms of the edge magnetic moments. Circles represent the amplitude of the Wannier functions 
 corresponding to two of these edge states on the nanoribbon sites (one shown on the upper, and one on the lower edge). 
}
\label{fig1}
\end{figure}

Returning to  pure zigzag nanoribbons, 
it was recently shown~\cite{Koop15} that  similarly to the chiral case, effective quantum spin models with a two-leg ladder geometry can also be derived for the case of wide zigzag nanoribbons, starting from the Hubbard model description, cf. the inset of Fig.~\ref{fig1} for an illustration. In contrast to the case of the chiral nanoribbons, however these effective spin models have  not  been further  analyzed  with respect to their magnetic properties. 


%
Here, we   study these effective  spin-ladder  models for zigzag nanoribbons using large-scale quantum Monte Carlo simulations~\cite{Sandvik03,Fukui09,Golor14}. This allows us to  account within the effective quantum spin model for  quantum fluctuations beyond mean-field-theory, while we can also access the large-distance correlations. 
As will be discussed in detail  below, the effective spin models for the zigzag nanoribbons exhibit
a relatively weak spatial decay of the intra-edge spin-spin interactions. More specifically,  as a function of the lateral distance $r$ between two spins, the numerically determined values of the ferromagnetic intra-edge exchange interactions fit well to a power-law asymptotic decay proportional to $1/r^2$,  while the antiferromagnetic inter-edge interactions decay faster, approximately 
proportional to $ 1/r^4$  at large values of $r$. 
This
results in a qualitatively different magnetic behavior as compared to the chiral case~\cite{Golor14}: we find that the quantum disordered region, which characterized the ground state of the effective quantum spin model for chiral ribbons, is reached in the case of the zigzag effective spin model only  upon further increasing the  antiferromagnetic inter-edge coupling strength beyond a finite critical value, which defines a quantum critical point at a rather large value of the inter-edge coupling strength. 

We determine the  critical scaling properties at this quantum critical point explicitly for a  simplified version of the effective spin model, wherein the antiferromagnetic inter-edge coupling is truncated beyond its nearest-neighbor  term.  In fact, this more genuine quantum spin model can  be seen as a basic spin model of two antiferromagnetically coupled ferromagnetic Haldane-Shastry  spin-half chains~\cite{Haldane88, Shastry88, Haldane91}. A single ferromagnetic Haldane-Shastry  chain has a ferromagnetic ground state and its thermodynamic properties have been obtained within a well-known exact  solution~\cite{Haldane91}. In this paper we show that 
the system of two antiferromagnetically coupled  Haldane-Shastry chains  features a  quantum phase transition between a low-coupling gapless phase and a strong-coupling quantum disordered region where dominant singlets form along the inter-chain bonds. 
We determine numerically the critical properties of the quantum critical point that separates these two phases and compare our estimates for the critical scaling exponents to recent predictions based  on renormalization group (RG) calculations performed in the context of critical O(3) $\phi^4$-theories, quantum rotor models and quantum non-linear sigma models with power-law interactions~\cite{Dutta01, Maghrebi16, Defenu17}. We observe good overall agreement between our numerically extracted values for the critical exponents and the RG findings, adding further support to the identification of the quantum phase transition in the effective spin model from identifying its universal properties.  

The  outline of the rest of this paper is as follows: in Sec.~\ref{sec:model}, we define in more detail the effective quantum spin model that we consider in our analysis, as well  as the quantum Monte Carlo approach that we use. We present our results for the phase diagram and the properties of the quantum critical point in Sec.~\ref{sec:results}, and finally provide a discussion of our numerical findings and the relation to graphene zigzag nanoribbons in Sec.~\ref{sec:discussion}.

\section{Model and Method}\label{sec:model}
In the following, we consider the effective quantum spin model for zigzag nanoribbons derived in Ref.~\onlinecite{Schmidt13, Koop15}, 
 which maps onto a spin-half Heisenberg model on a two-leg ladder, described by the Hamiltonian
\begin{equation}
H=-\sum_{i,j} J^\mathrm{F}_{ij}\:  (\mathbf{S}_{i,1}\cdot \mathbf{S}_{j,1} +\mathbf{S}_{i,2}\cdot \mathbf{S}_{j,2}) + 
\sum_{i,j} J^\mathrm{AF}_{ij}  \: \mathbf{S}_{i,1}\cdot \mathbf{S}_{j,2},
\end{equation}
where $\mathbf{S}_{i,\mu}$ denotes a spin on the $i$-th rung of the two-leg ladder, which for $\mu=1$ ($2$) is located on the upper (lower) leg.  
Furthermore, $J^\mathrm{F}_{ij}>0$ denotes the magnitude of the ferromagnetic exchange interaction for spins located on the same leg, and 
$J^\mathrm{AF}_{ij}>0$  is the antiferromagnetic coupling between  spins on different legs. Due to translational symmetry, these couplings depend only on the
lateral distance $r_{ij}=|i-j|$, i.e.,  $J^\mathrm{(A)F}_{ij}=J^\mathrm{(A)F}(r_{ij})$. 
The actual values of the coupling constants, obtained
as described in Refs.~\onlinecite{Schmidt13, Koop15}, depend explicitly on the physical parameters of the zigzag nanoribbon within the Hubbard model description.
To leading order, the ferromagnetic couplings scale proportional to the local Hubbard repulsion $U$, and the antiferromagnetic couplings scale with $t^2/U$, where $t$ denotes the nearest-neighbor hopping strength. 
For concreteness,  we consider here the case where $U=t$, well within the semi-metallic region for the Hubbard model on a  honeycomb lattice, in accord with the conditions in bulk graphene~\cite{CastroNeto09}. In the following, we consider a zigzag nanoribbon that is sufficiently wide, such that 
the edge magnetic moments are well described within  the Wannier function basis~\cite{Schmidt13, Koop15}.
We thus choose explicitly a zigzag nanoribbon with $W=10$ zigzag lines (cf. the inset of Fig.~\ref{fig1} for an illustration), for which we obtained the effective coupling strengths  given in the main panel of Fig.~\ref{fig1} as well as, for $r\leq 4$, in  Tab.~\ref{tab:c}. These are based on a calculation for a $W=10$ nanoribbon with 48000 lattice sites. Furthermore, from comparing the results for $W=10$ zigzag nanoribbons of varying sizes, we ensured that the shown values of the effective couplings are not affected by finite-size effects. 

The  log-log plot Fig.~\ref{fig1}  exhibits an essentially algebraic decay of the calculated exchange couplings as a function of distance for values of $r \gtrsim 5$, traced over several  orders of magnitude in the interaction strength. As indicated by the corresponding fit lines, this large-$r$ behavior  is captured reasonably well in terms of  asymptotic algebraic decays $J^\mathrm{F}(r)\propto 1/r^2$ and  $J^\mathrm{AF}(r)\propto 1/r^4$, respectively.  Fitting the couplings to power-law decays, one obtains estimated exponents of $1.94$ and $4.09$, respectively. Given the approximative nature of the coupling constant calculation, we prefer to employ  for the further analysis  the very close, and more natural values of 2 and 4, respectively.
 We considered also other values of the nanoribbon width, $W= 8$ and $12$, and also for these ribbons   the above asymptotic algebraic decays  do fit  the numerical results similarly well . 
Based on these algebraic forms,  we can thus use 
the following  explicit form of the longer-ranged coupling constants in the Hamiltonian $H$: 
\begin{equation}\label{eq:coupfits}
 J^\mathrm{F}(r)=J_\mathrm{F}\:\frac{1}{r^2},\quad J^\mathrm{AF}(r)=J_\mathrm{AF}\:\frac{1}{r^4}, \quad r>4,
 \end{equation}
 with the fit parameters $J_\mathrm{F}/t=0.01009$ and $J_\mathrm{AF}/t=0.21972$,
 while for smaller distances,  the  values of the interactions for $W=10$ are given explicitly in Tab.~\ref{tab:c}.

\begin{table}[t]
\begin{center}
\begin{tabular}{| c | c | c |}
\hline
$r$ & $J^\mathrm{F}(r)/t$ & $J^\mathrm{AF}(r)/t$ \\
\hline
0 &       --            & 0.0196417 \\
1 & 0.0453914 &  0.0155852 \\
2 & 0.0047475 & 0.0079814 \\
3 & 0.0014386 & 0.0028894 \\
4 & 0.0007365 & 0.0008942\\
\hline
\end{tabular}
\end{center}
\caption{\label{tab:c} Values of the  spin exchange couplings for lateral distances $r \leq 4$, as obtained for the effective spin-ladder model for the $W=10$ nanoribbon for $U=t$. 
}
\end{table}

%

In order to  systematically study the physics of the Hamiltonian $H$ with the above  form of the couplings, it turns out instructive to
tune the relative strength of the inter-leg to intra-leg couplings beyond these original values. For this purpose, we introduce 
a dimensionless quantity $\lambda$, which uniformly  rescales all the inter-leg couplings  as
\begin{equation}
J^\mathrm{AF}(r) \rightarrow \lambda \: J^\mathrm{AF}(r).
\end{equation} 
Hence, for $\lambda =1$ we recover the original model, while for larger $\lambda$ we (artificially) enhance all  antiferromagnetic inter-leg couplings with respect to the ferromagnetic intra-leg couplings.
As will be demonstrated in the next section, the Hamiltonian $H$ indeed exhibits a quantum phase transition upon varying the  parameter $\lambda$, which we referred to already.

Furthermore, we find that the basic physics of the Hamiltonian $H$ is reproduced also for a simplified model Hamiltonian, which is obtained by truncating the antiferromagnetic exchange couplings beyond the nearest-neighbor term and using a simple $1/r^2$ decay for all ferromagnetic couplings $r\geq 1$. This leads us to an even more genuine spin model with Hamiltonian 
\begin{equation}
\tilde{H}=
-J_{F} \sum_{i,j} \frac{1}{r_{ij}^2}  (\mathbf{S}_{i,1}\cdot \mathbf{S}_{j,1} +\mathbf{S}_{i,2}\cdot \mathbf{S}_{j,2}) + 
J_\mathrm{AF}  \sum_{i}  \: \mathbf{S}_{i,1}\cdot \mathbf{S}_{i,2}.
\end{equation}
For this model, we furthermore define the ratio 
\begin{equation}
g=\frac{J_\mathrm{AF}}{J_\mathrm{F}}
\end{equation}
between the two coupling parameters. Similarly to the  parameter $\lambda$ in the Hamiltonian $H$, $g$ quantifies for the Hamiltonian $\tilde{H}$ the relative strength of the antiferromagnetic inter-leg coupling with respect to the ferromagnetic intra-leg coupling strength.  

In the limit of $J_\mathrm{AF} =0$ (i.e., $g=0$), this model corresponds to two decoupled spin chains with a ferromagnetic $1/r^2$ exchange interaction. In the thermodynamic limit, this is the ferromagnetic Haldane-Shastry model, for which an exact solution has been derived for its thermodynamic properties~\cite{Haldane88,Shastry88,Haldane91}. This model has a fully polarized, ferromagnetic ground state.
Given  the short-ranged character of the inter-leg coupling in $\tilde{H}$, we 
expect in this case a quantum disordered phase from the formation of strong rung-singlets in the opposite limit of large $J_{AF}$, i.e., for  $g\rightarrow \infty$,
along with a finite spin excitation gap.  
Note that due to the explicit ferromagnetic $1/r^2$-coupling between any two spins within a given leg, the correlation function decays proportional to $1/r^2$ even
deep inside the quantum disordered region, as one also  finds explicitly within perturbation theory about the large-$g$ limit. 

Any finite value  of the antiferromagnetic rung coupling, $g>0$, tends to lock the spins between the two legs into an antiferromagnetic alignment. However,
in contrast to the case of a purely short-ranged intra-leg coupling~\cite{Roji96,Kolezhuk96,Vekua03}, this locking does not immediately destroy the ferromagnetic state along each leg, due to the long-ranged character of the intra-leg coupling. 
Instead, as demonstrated in the next section, a quantum phase transition emerges at a finite value of $g>0$, which separates the weak coupling (low-$g$) from the 
strong coupling (large-$g$) phase. In this sense, the long-ranged character of the ferromagnetic  intra-leg coupling stabilizes the weak coupling  phase, 
in contrast to the case of the conventional two-leg ladder, where any finite rung coupling drives the system into the gapped rung-singlet regime. 

Since for the Hamiltonian $H$, the antiferromagnetic  inter-leg coupling decays  fast with the lateral distance (as compared to the intra-leg couplings), this extended form of the inter-leg coupling
does not modify the above  picture. In fact, as we will show in the following section, also for $H$ we  can identify a quantum critical point at a finite value of $\lambda$. 
One may indeed expect this, given the fact that quite generally in one-dimensional systems, power-law interactions decaying faster than $1/r^3$ lead to 
the same critical properties as  short-ranged interactions. 

Before we turn to the presentation of our results, we comment on the numerical approach that we used for our investigation. 
We  analyzed the  properties of the model Hamiltonians $H$ and $\tilde{H}$ using quantum Monte Carlo (QMC) simulations. In fact,  both models are free of geometric frustration, so that no QMC sign-problem occurs. To efficiently perform the QMC sampling in the presence of the long-ranged interactions, we used 
the stochastic series expansion QMC  method for quantum spin systems with an efficient sampling scheme~\cite{Sandvik03,Fukui09}, similar as in previous studies for the effective spin model for chiral nanoribbons~\cite{Golor14}. In particular, we simulated finite two-leg ladder systems with the Hamiltonians $H$ and $\tilde{H}$  using periodic boundary conditions (PBC) along the lateral direction. 
In order to reduce finite-size effects in the QMC simulations and  access more efficiently the behavior of the effective spin models on large distances, we furthermore performed an Ewald summation of the long-ranged effective spin interactions~\cite{Fukui09}. For a given pair of spins with lateral distance $r$, we thus replace the coupling constant for the finite system with $L$ rungs (i.e., $L$ spins on each leg of the two-leg ladder, and a total of $N=2L$ spins) by 
a summation over all replica-repeated images.
In particular, for the ferromagnetic couplings in  $\tilde{H}$, we obtain a closed form, since the Ewald summation leads to 
\begin{equation}
\frac{J_\mathrm{F}}{r^2} \longrightarrow\sum_{k=-\infty}^\infty \frac{ J_\mathrm{F}}{(r-k L)^2}= J_\mathrm{F} \frac{ {\pi^2}/{L^2} }{\sin^2\left(\frac{r\: \pi}{L}\right)}=\frac{J_\mathrm{F}}{\zeta(r)^2},
\end{equation}
where the closed form  of the above series can be found, e.g.,  in Ref.~\onlinecite{Fukui09},
and we defined
\begin{equation}
\zeta(r)= \sin(r\: \pi/L) L / \pi.
\end{equation} 
One may notice that the above closed form of the ferromagnetic coupling for PBC is also usually considered in the  Haldane-Shastry model for finite chains, and in fact, 
$\zeta(r_{ij})= \sin(r_{ij}\: \pi/L) L / \pi$ equals the chord distance for a periodic chain with $L$ sites between lattice sites $i$ and $j$.  In the context of conformal field theory, $\zeta$ is  often called the conformal distance (or length), and we will also use this notation  further below.
For the  couplings in the Hamiltonian $H$ we  also  performed a corresponding Ewald summation,
for which we however do not obtain a closed form, and instead 
performed the summation numerically. 

In the following section, we present our results from QMC simulations of both model Hamiltonians. Since our QMC method is a finite temperature scheme, we monitor the behavior of various physical quantities in the low-temperature region in order to extract the ground state behavior, considering system sizes with typically $20.000$ and in some cases up to $32.000$ quantum spin-half sites. 
Furthermore, we use 
units in the following such that the nearest neighbor ferromagnetic intra-leg coupling is set equal to one, i.e., $J^\mathrm{F}(r=1)=1$ for $H$, and  $J_\mathrm{F}=1$ for $\tilde{H}$, 
respectively. In addition we use $k_B=1$.

\section{Results}\label{sec:results}

In the following subsection, we show that both models $H$ and $\tilde{H}$ feature a quantum phase transition between a gapless phase at weak inter-leg coupling and the gapped, quantum disordered phase for strong inter-leg couplings. In the next subsection, we then analyze the scaling behavior at the quantum critical point,  focusing  on  the more genuine Hamiltonian $\tilde{H}$ and compare our results to recent RG calculations on related quantum systems. 

\subsection{Quantum phase transition}\label{sec:A}
In the absence of any inter-leg coupling, both models consist of two decoupled ferromagnetic Haldane-Shastry chains~\cite{Haldane88,Shastry88,Haldane91}, and each chain has a long-ranged ordered, ferromagnetic ground state. At any finite temperature $T$, this ferromagnetic order is destroyed, with a correlation length that increases exponentially upon decreasing $T$. Correspondingly, an isolated ferromagnetic Haldane-Shastry chain exhibits 
an exponential divergence  of the magnetic susceptibility~\cite{Haldane91} upon lowering $T$. 
In order to probe
the low temperature behavior of the magnetic correlations within each leg of the coupled-chain systems, 
 in the QMC simulations we measured the corresponding single-leg susceptilbility 
\begin{equation}\label{eq:chilegdef}
\chi_\mathrm{leg}=\frac{1}{L} \int_0^{1/T} d\tau \: \langle M_\mathrm{leg} (\tau)  M_\mathrm{leg} (0)\rangle,
\end{equation}
where $M_\mathrm{leg}=\sum_i S^z_{i,\mu}$ denotes the total magnetic moment of the spins on one of the legs, and where $\mu=1$ or $\mu=2$ can be chosen equally well (within the QMC simulations, we average over both cases in order  to improve the statistics). The above Kubo-integral  quantifies the fluctuations of the single leg's magnetic moment, with $\tau$ denoting the imaginary-time evolution. 
Here, we employ the SU(2)-symmetry of the quantum spin Hamiltonian in order to evaluate the magnetic correlations directly in the computational $(S^z)$ basis. 
Physically,  $\chi_\mathrm{leg}$ quantifies the linear response in the leg's magnetic moment $M_\mathrm{leg}$ upon applying a uniform magnetic field along a single leg of the ladder. 

The overall 
magnetic response of the two-leg ladder models is obtained from the uniform susceptibility 
\begin{equation}
\chi_\mathrm{uni}=\frac{1}{N} \int_0^{1/T} d\tau \: \langle M(\tau) M(0)\rangle= \frac{1}{T N}\langle M^2 \rangle, 
\end{equation} 
in terms of the   fluctuations in the  total system's ($N=2L$) magnetic moment $M=\sum_{i,\mu} S^z_{i,\mu}$. Note that while $M$ commutes with both Hamiltonians,  this is not the case for  $M_\mathrm{leg}$ at any finite inter-leg coupling. In physical terms,  $\chi_\mathrm{uni}$ quantifies the linear response in the total magnetic moment $M$ upon applying a uniform magnetic field to all the spins of the system.

In addition to the above quantities, one may also consider the overall system's staggered susceptibility 
\begin{equation}
\chi_\mathrm{stag}=\frac{1}{N} \int_0^{1/T} d\tau \: \langle M_\mathrm{stag} (\tau)  M_\mathrm{stag} (0)\rangle,
\end{equation}
where $M_\mathrm{stag}=\sum_i (S^z_{i,1}-S^z_{i,2})$.
However,  
since $\chi_\mathrm{stag}=2\chi_\mathrm{leg}-\chi_\mathrm{uni}$, and (as we will  also find from explicit calculations) $\chi_\mathrm{leg} \gg \chi_\mathrm{uni}$ at low temperatures due to the antiferromagnetic inter-leg coupling,
$\chi_\mathrm{stagg}$ essentially probes the intra-leg ferromagnetic response, which $\chi_\mathrm{leg}$ accesses more directly. From the point of view of the edge-magnetism, 
$\chi_\mathrm{leg}$, probing for ferromagnetic correlations within a single leg,  
may also appear  to be the more natural quantity to consider.

\begin{figure}[t]
\includegraphics[width=\columnwidth]{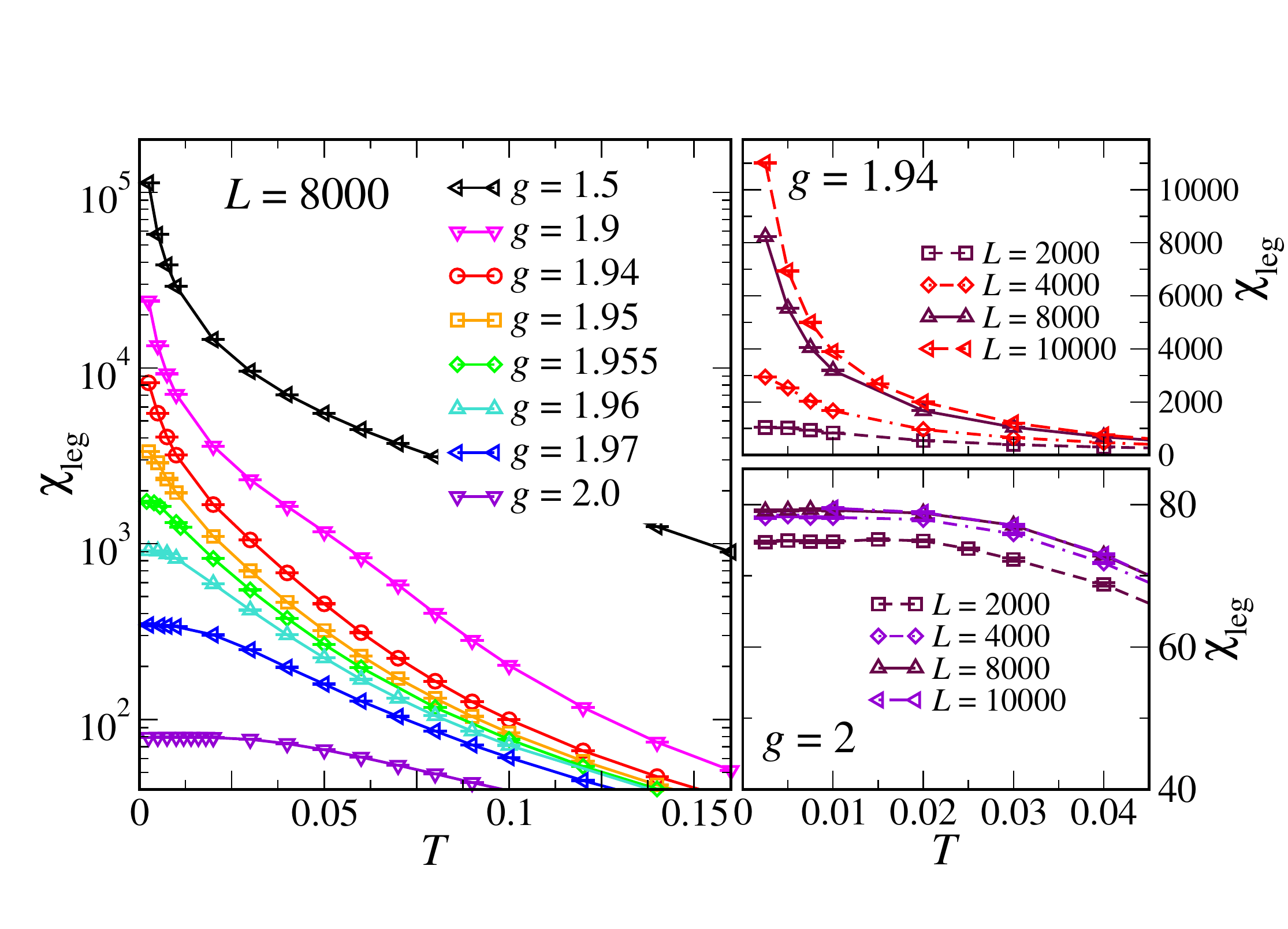}
\caption{(Color online) Temperature dependence of the single-leg susceptibility $\chi_\mathrm{leg}$ of the Hamiltonian $\tilde{H}$ for different values of $g$ and for $L=8000$  (left panel), and for different values of $L$ at fixed $g=1.94$ (upper right panel), and $g=2$ (lower right panel).}
\label{fig2}
\end{figure}

We first consider the evolution of these quantities upon varying the parameter $g$ for the Hamiltonian $\tilde{H}$. The left panel of Fig.~\ref{fig2}
shows the low temperature behavior of the single-leg susceptibility $\chi_\mathrm{leg}$ for a system with $L=8000$ and for different values of $g$ in a region, where we observe a strong qualitative change in the low-$T$ behavior. Namely, for values of $g<1.95$, the single-leg susceptibility develops a strong divergence upon lowering $T$, similar to the case of an isolated ferromagnetic Haldane-Shastry model, while for values of $g>1.96$, this divergence is suppressed at low temperatures, and $\chi_\mathrm{leg}$ instead tends to a finite value for $T\rightarrow 0$. 
An exponential divergence of the magnetic susceptibility, such as obtained
for a  single ferromagnetic Haldane-Shastry chain, 
is affected by finite-size effects in QMC simulations~\cite{Vassiliev01}, so that in the low-temperature region, one needs to carefully monitor  the behavior of the susceptibility upon varying the system size. This is shown for the two cases of $g=1.94$ and $g=2$ in the two right panels of  Fig.~\ref{fig2}. We find that  for $g=2$, the finite-size data shows a convergent saturation in the low-temperature value of $\chi_\mathrm{leg}$, while the data for $g=1.94$ shows a steady increase upon increasing the system size. This rather drastic change in the low-temperature behavior of the single-leg response function upon a weak variation of the coupling ratio $g$ by only a few percent  is indicative of a quantum phase transition of the model within this parameter region. 

\begin{figure}[t]
\includegraphics[width=\columnwidth]{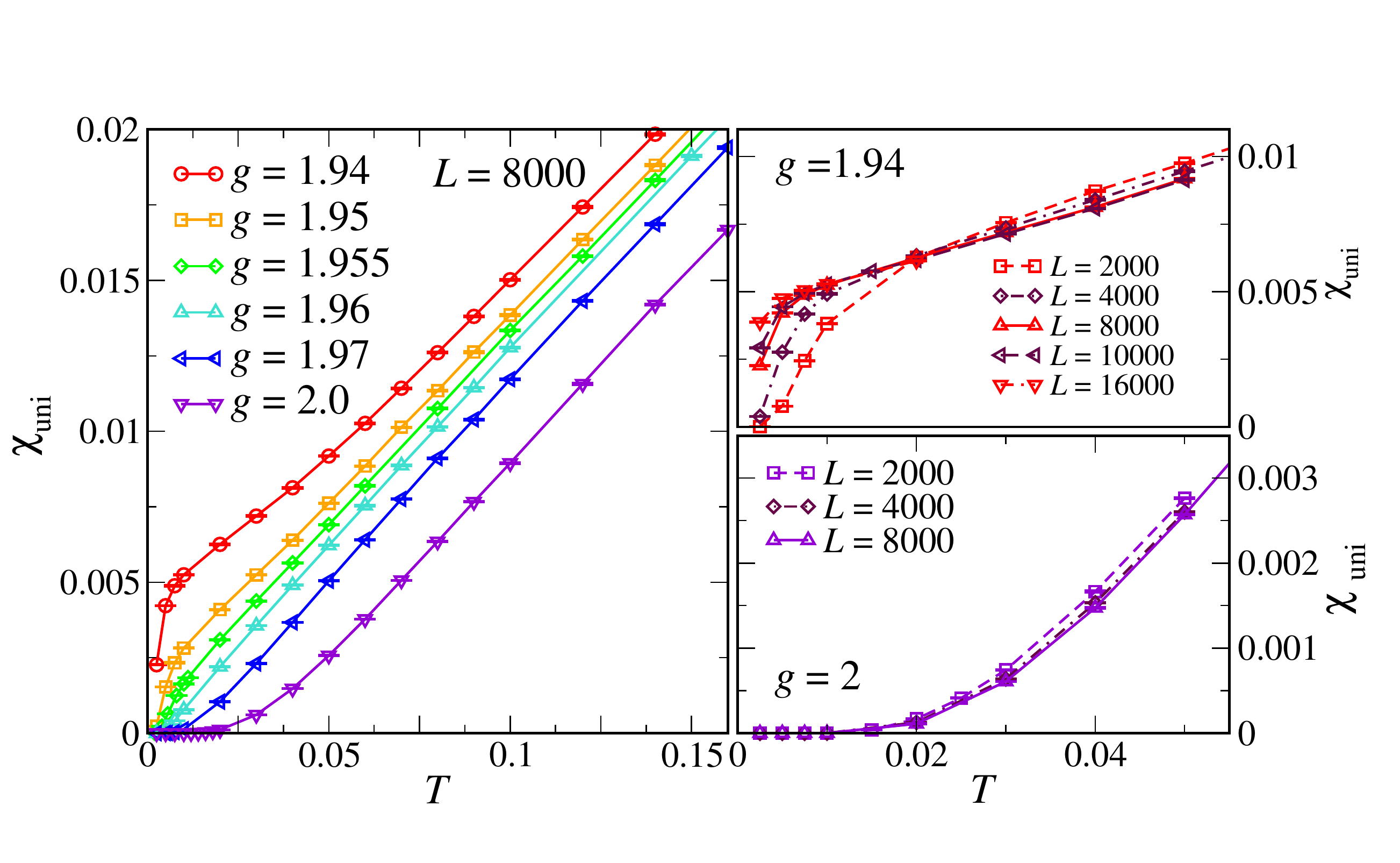}
\caption{(Color online)
Temperature dependence of the uniform susceptibility $\chi_\mathrm{uni}$ of the Hamiltonian $\tilde{H}$ for different values of $g$ and for $L=8000$  (left panel), and for different values of $L$ at fixed $g=1.94$ (upper right panel), and $g=2$ (lower right panel).}
\label{fig3}
\end{figure}

We obtain further indication for a change in the ground state properties  from analyzing the uniform magnetic susceptibility $\chi_\mathrm{uni}$, for which our QMC results are shown in Fig.~\ref{fig3}. From the temperature dependence of  $\chi_\mathrm{uni}$, shown in the left panel of Fig.~\ref{fig3}, we find  for values of  $g<1.95$ a leading linear  behavior that extrapolates to finite  ground state values. The sudden drop of $\chi_\mathrm{uni}$  that sets in at very low temperatures is in fact a finite-size effect, as can be seen from a detailed view of the low-$T$ behavior of  $\chi_\mathrm{uni}$ for different system sizes in the upper right panel of Fig.~\ref{fig3} for $g=1.94$. In contrast, for $g>1.96$, the low-temperature data shows a strong suppression of the magnetic response  $\chi_\mathrm{uni}$, which becomes  more pronounced upon increasing the system size, cf. the lower right panel for $g=2$. The finite-size effects for $g=1.94$ (upper right panel) can also be distinguished from the low-$T$ suppression of $\chi_\mathrm{uni}$ for $g=2$ (lower right panel) by a different curvature in the temperature dependence. Hence, similarly to the single-leg susceptibility, the uniform susceptibility exhibits a strong, qualitative change in the system's behavior in the vicinity of $g\approx 1.955$. Moreover, the vanishing uniform susceptibility for $g\gtrsim 1.955$ indicates the presence of a finite spin excitation gap $\Delta$. In the next subsection, we will quantify the spin gap by extracting it from the low-temperature data for $\chi_\mathrm{uni}$, and  also compare its dependence on the coupling ratio to  predictions from scaling theory. 

\begin{figure}[t]
\includegraphics[width=\columnwidth]{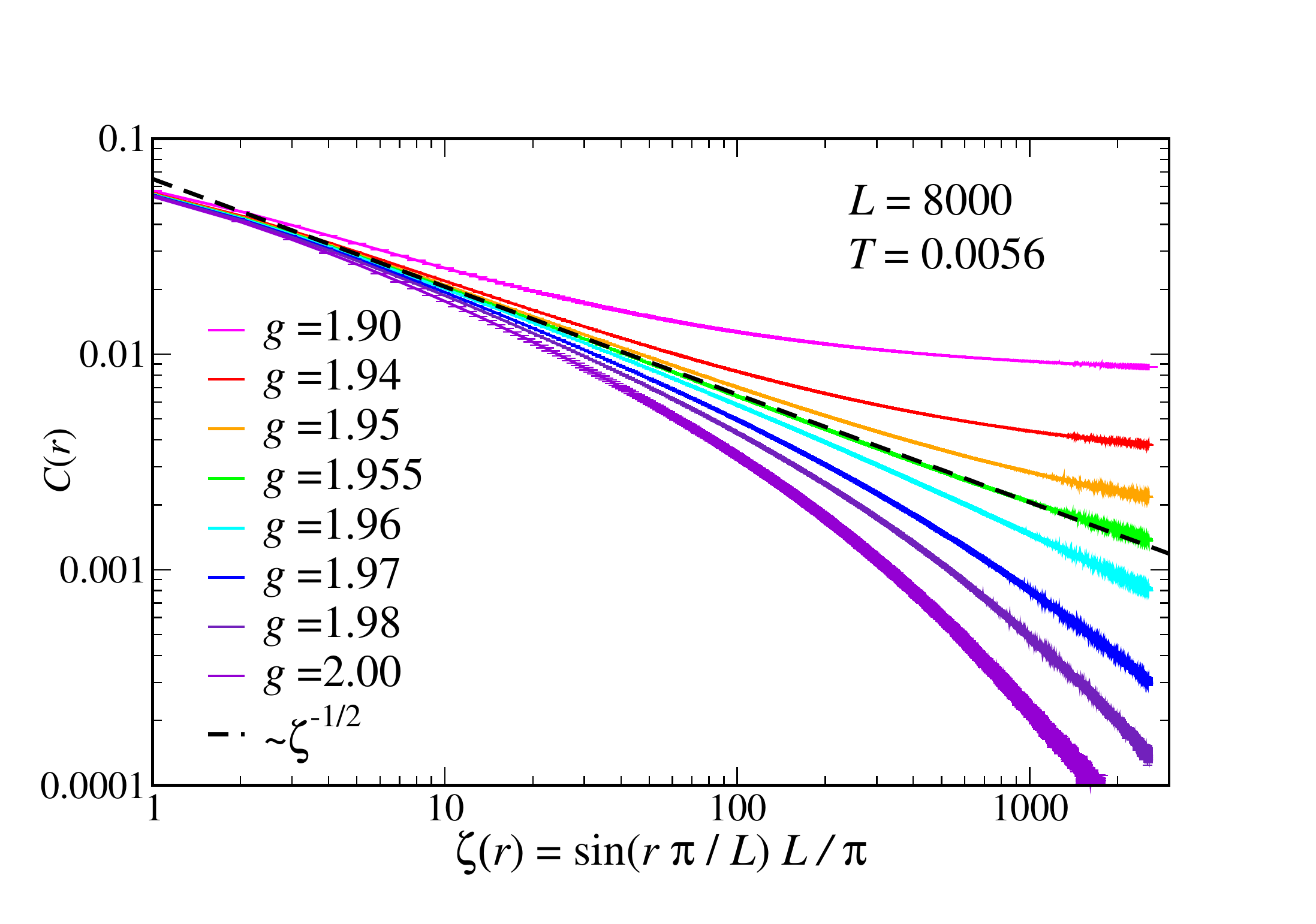}
\caption{(Color online) Intra-leg spin correlations $C(r)$ as a function of the conformal distance $\zeta(r)=\sin(r\: \pi /L) L /\pi$ for the Hamiltonian $\tilde{H}$ and for different values of $g$ and for $L=8000$ at $T=0.0056$. The dashed line indicates a scaling proportional to $\zeta^{-1/2}$ near the quantum critical point. }
\label{fig4}
\end{figure}

The above analysis of the thermodynamic response functions gives strong indication for the presence of a quantum phase transition in the system described by $\tilde{H}$. In order to relate this observation more directly to the spin correlations within the legs of the coupled two-leg ladder system, we examine  the correlation function 
\begin{equation}
C(r_{ij})=\langle S^z_{i,\mu} S^z_{j,\mu} \rangle
\end{equation}
within a single leg ($\mu=1$ or $2$), which is shown as obtained from QMC simulations on an $L=8000$ system at a low temperature of $T=0.0056$, and for different values of $g$ within the transition region in Fig.~\ref{fig4}. Here, we furthermore use the conformal distance $\zeta(r)=\sin(r\: \pi /L) L /\pi$ to quantify the lateral separation between the spins. We find again a qualitative change of the large-distance behavior of $C(r)$ at $g\approx 1.955$. For smaller values of $g$, the correlation function has a different curvature than the data for $g>1.955$, which 
furthermore shows a strong suppression at large $r$. Moreover, the data for $g=1.955$ compares well to an algebraic decay  proportional to $\zeta^{-1/2}$, indicated by the dashed line in Fig.~\ref{fig4}. Such an algebraic scaling behavior of the finite-system's correlation function in terms of the conformal distance is characteristic for the decay of the correlation function at a quantum critical point, with an emerging conformal invariance in $1+1$-dimensional quantum systems. Note also that this slow algebraic decay is distinct from the asymptotic $1/r^2$-decay in the large-$g$ region, which stems from the explicit ferromagnetic couplings decaying as $1/r^2$.

Summarizing  these  results, we have obtained indication from both two-point correlation functions and global quantities that the Hamiltonian $\tilde{H}$ exhibits a quantum phase transition at $g\approx 1.955$ between a low-$g$ gapless phase with long-ranged ferromagnetic correlations along each leg, and a large-$g$ quantum disordered region with a finite spin excitation gap. Furthermore, an approximately algebraic decay of the correlation function is indicative of a quantum critical point, separating the two different phases. In the following subsection, we will confirm this basic observation by studying   the properties of this quantum critical point within a more detailed finite-size scaling analysis. 

\begin{figure}[t]
\includegraphics[width=\columnwidth]{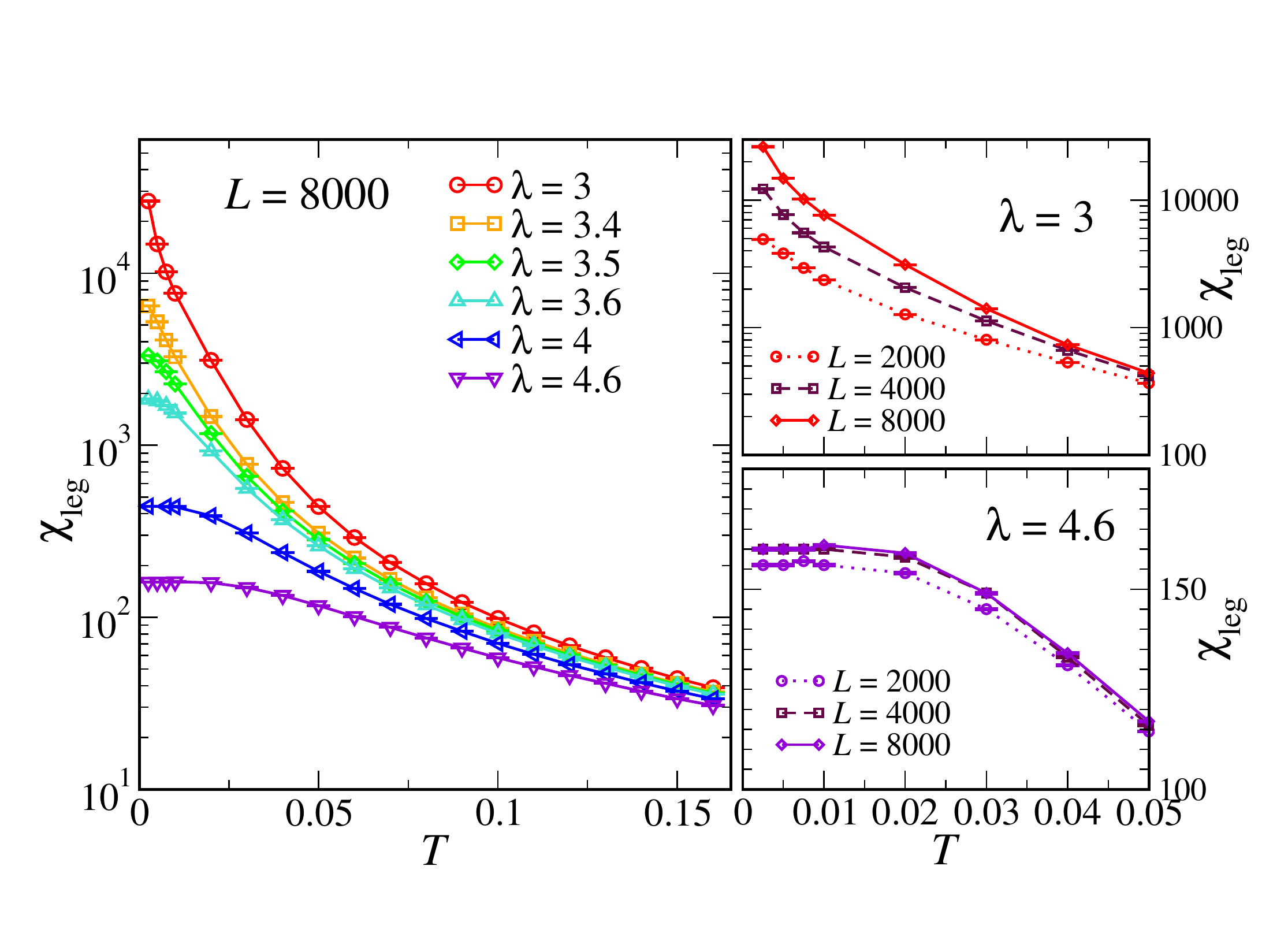}
\caption{(Color online) Temperature dependence of the single-leg susceptibility $\chi_\mathrm{leg}$ of the Hamiltonian ${H}$ for different values of $\lambda$ and for $L=8000$  (left panel), and for different values of $L$ at fixed $\lambda=3$ (upper right panel), and $\lambda=4.6$ (lower right panel).}
\label{fig5}
\end{figure}

\begin{figure}[t]
\includegraphics[width=\columnwidth]{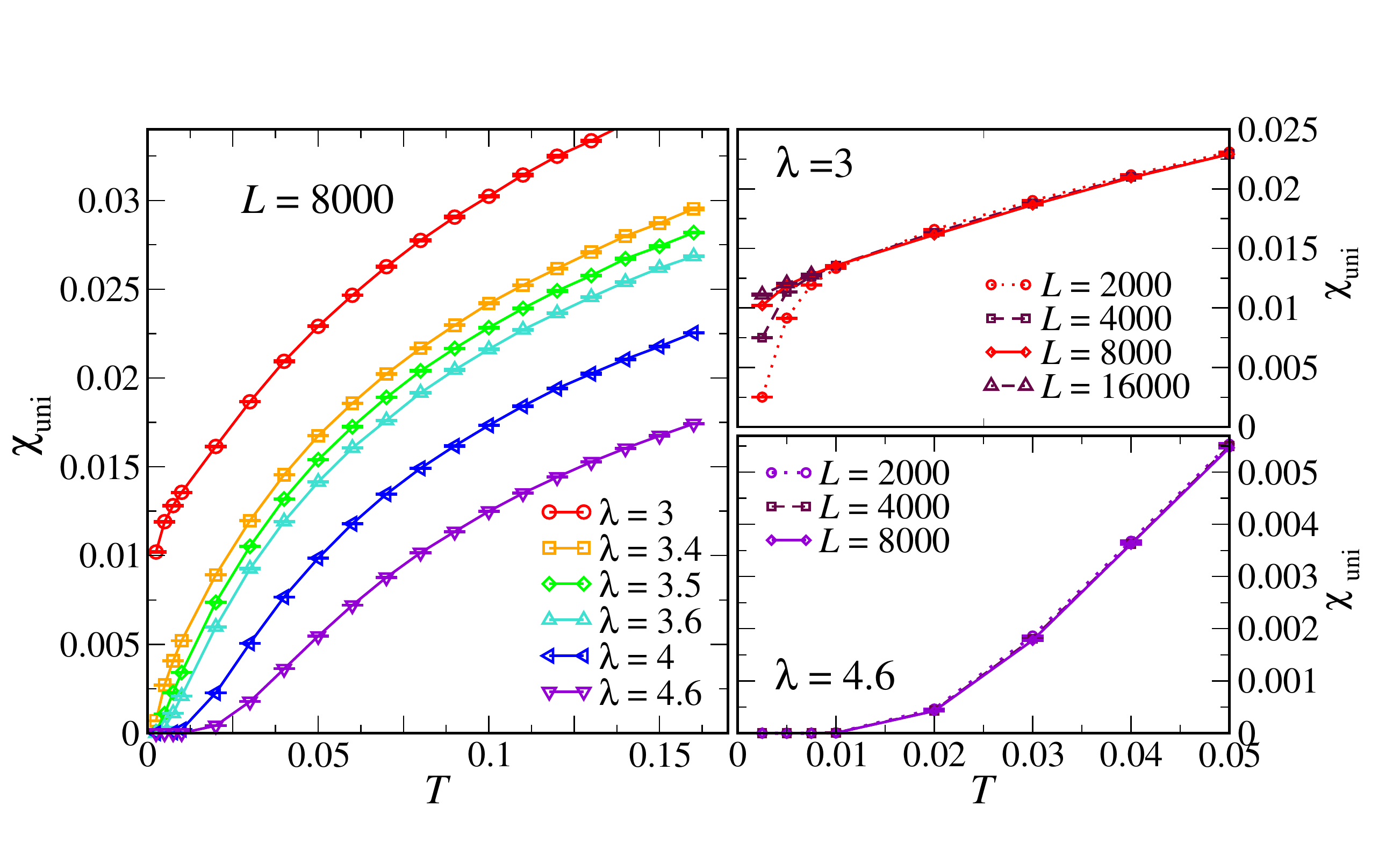}
\caption{(Color online)
Temperature dependence of the uniform susceptibility $\chi_\mathrm{uni}$ of the Hamiltonian ${H}$ for different values of $\lambda$ and for $L=8000$  (left panel), and for different values of $L$ at fixed $\lambda=2.05$ (upper right panel), and $\lambda=2.1$ (lower right panel).}
\label{fig6}
\end{figure}

\begin{figure}[t]
\includegraphics[width=\columnwidth]{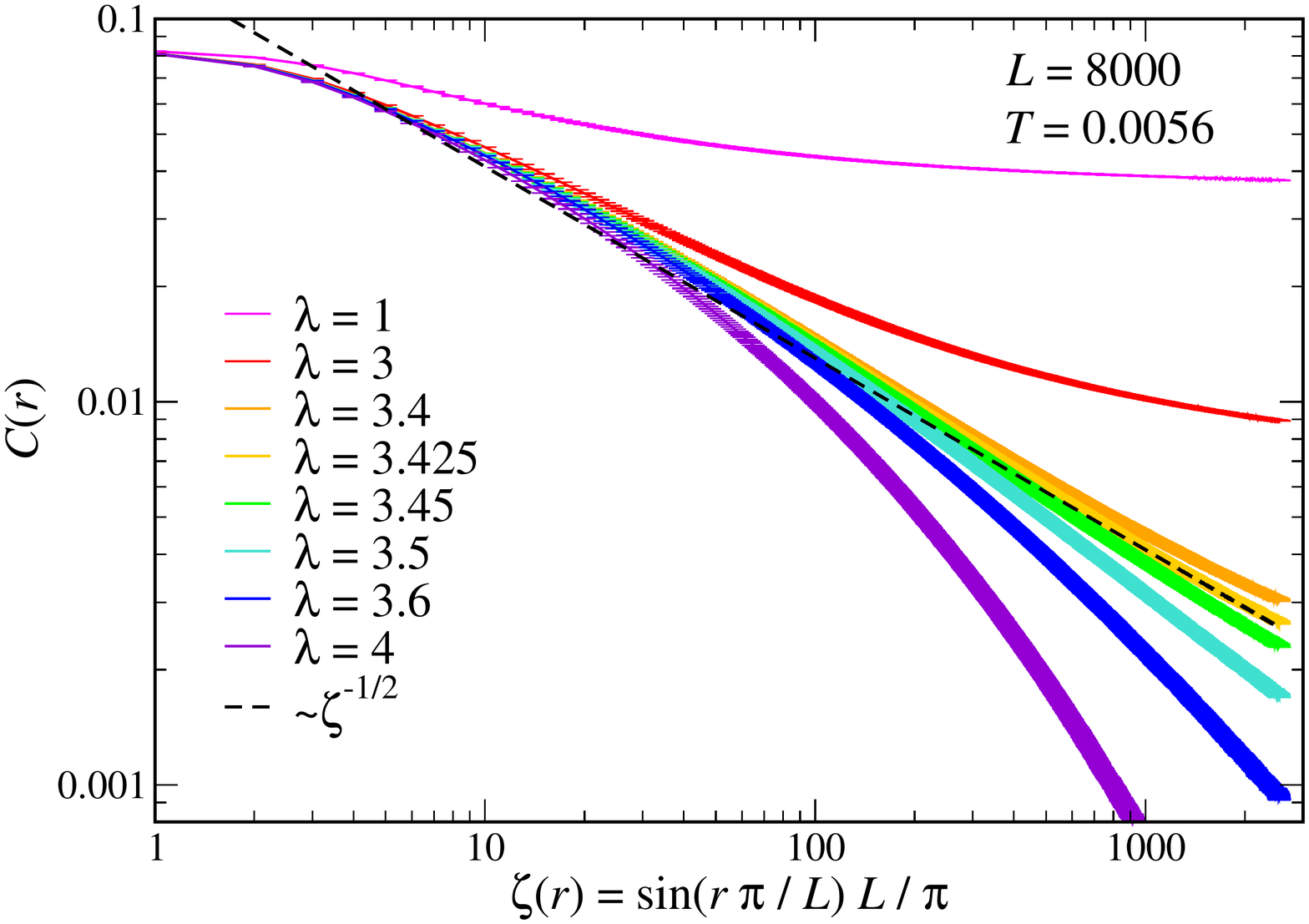}
\caption{(Color online) Intra-leg spin correlations $C(r)$ as a function of the conformal distance $\zeta(r)=\sin(r \: \pi / L) L /\pi$ for the Hamiltonian ${H}$ and for different values of $\lambda$ and for $L=8000$ at $T=0.0056$. The dashed line indicates a scaling proportional to $\zeta^{-1/2}$ near the quantum critical point. }
\label{fig7}
\end{figure}

\begin{figure}[t]
\includegraphics[width=\columnwidth]{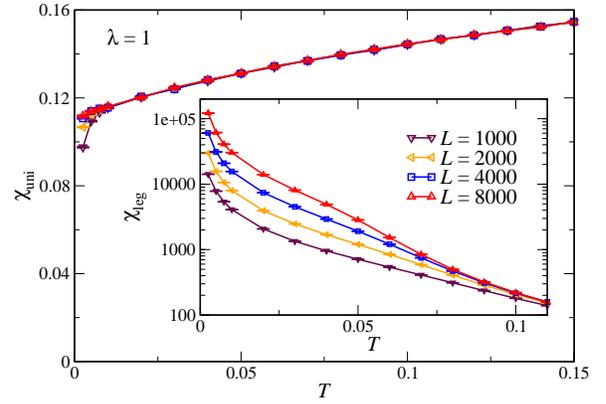}
\caption{(Color online) Temperature dependence of the 
uniform susceptibility $\chi_\mathrm{uni}$ (main panel) and the  
single-leg susceptibility $\chi_\mathrm{leg}$ (inset) of the  Hamiltonian ${H}$ for $\lambda=1$ for different values of $L$.}
\label{fig8}
\end{figure}

Before we turn to this scaling analysis of the quantum critical point, we show that a similar behavior is also obtained for the 
Hamiltonian $H$, for which the inter-leg interactions have an extended $1/r^4$-decay, instead of the nearest-neighbor inter-leg coupling in $\tilde{H}$. 
In Fig.~\ref{fig5} and Fig.~\ref{fig6}, we show our QMC results for $\chi_\mathrm{leg}$ and $\chi_\mathrm{uni}$  for the model Hamiltonian $H$. Indeed, 
we find clear indication for a qualitative change in the system's properties upon increasing $\lambda$, and  we estimate a critical coupling ratio of $\lambda_c\approx 3.4$.
This value is  consistent with the large distance  behavior of the correlation function $C(r)$, shown in Fig.~\ref{fig7}, which  indicates a quantum critical point located at $\lambda_c\approx 3.425$. This plot also contains the  correlation function $C(r)$ for $\lambda=1$, and the corresponding data for the susceptibilities is shown in Fig.~\ref{fig8}.
The original effective spin-ladder model $H$ for $\lambda=1$ is thus localized well within the weak-coupling gapless phase with long-ranged ferromagnetic alignment along each leg stabilized  in the ground state. 

%
\subsection{Quantum critical properties}\label{sec:B}
To further examine the quantum phase transition in the effective quantum spin models, we analyze in this subsection its critical scaling properties, focusing for this purpose on the more genuine case of Hamiltonian $\tilde{H}$. It is convenient to  first summarize some of the main findings from several recent RG studies of the quantum critical properties of related one-dimensional quantum  systems with an O($3$) symmetry in the presence of  long-ranged interaction~\cite{Dutta01, Maghrebi16, Defenu17}. 
Some of these papers also consider the more general case of an  O($n$) symmetric interaction~\cite{Dutta01,Defenu17}, while Ref.~\onlinecite{Maghrebi16} focuses on the case of $n=1$, which is relevant, e.g.,  for the quantum Ising model. 

For a quantum system in 1+1 dimensions, with a spatially long-ranged interaction that decays proportional to $1/r^{1+\sigma}$ with the spatial distance $r$, such as an $n$-component quantum rotor model, the long-ranged nature of the interactions is important in order to stabilize a non-trivial transition. For example,  short-ranged interacting quantum rotor models do not exhibit  quantum phase transitions for $n>2$ in 1+1 dimensions~\cite{Dutta01,Sachdev11}. Of particular interest to the current discussion is the case $\sigma=1$, $n=3$. 
In the relevant region of $\sigma$ (for $2/3<\sigma<2$), the critical exponents at the quantum phase transition differ from the mean-field values due to the effects of quantum fluctuations, and have been approximately obtained as  expansions in $\epsilon=3\sigma/2-d$~\cite{Dutta01,Maghrebi16}. Here, $d$ denotes the spatial dimension, and  $3\sigma/2$ is indeed  the upper critical dimension. In particular,  within the $\epsilon$-expansion, Ref.~\onlinecite{Dutta01} obtains from a one-loop calculation the result
\begin{equation}\label{eq:nu}
{\nu}=\frac{1}{\sigma}+\frac{n+2}{n+8}\epsilon+ \mathcal{O}(\epsilon^2)
\end{equation}
for the critical exponent $\nu$, which characterizes the divergence of the order parameter correlation length.
Another recent work~\cite{Maghrebi16} reports
$
{1}/{\nu}=\sigma-\frac{\epsilon}{3}+ \mathcal{O}(\epsilon^2)
$
for the special  case of $n=1$. This is in accord with the above result for  $\sigma=1$, the case of interest here,  but differs from it in the  general case (cf. Ref.~\onlinecite{Maghrebi16} for further discussion).  From Eq.~(\ref{eq:nu}), we thus  obtain an estimate of $\nu\approx 1.227$ for $\sigma=1$ and $n=3$. 
Ref.~\onlinecite{Maghrebi16} furthermore 
reports a two-loop order result for the 
dynamical critical exponent $z$ for $\sigma=1$, $d<2$ in  the O($n$) case, 
\begin{equation}
z=\frac{1}{2}+\frac{(n+2)(12-\pi^2)}{16(n+8)^2}\tilde{\epsilon}^2+\mathcal{O}(\tilde\epsilon^3),
\end{equation}
where $\tilde\epsilon=2-d$, based on earlier RG calculations~\cite{Pankov04}. For the case of interest here ($d=1$ and $n=3$), this provides an estimate of  $z\approx 0.505$. This value is also consistent with the RG results  in Ref.~\onlinecite{Defenu17}.
It may  be worthwhile to point out that within mean-field theory, a value of $z_\mathrm{MF}=\sigma/2<1$  results for $\sigma<2$, already reflecting the fact that due to the long-ranged nature of the spatial  interactions,  correlations in the temporal direction are weaker than in the spatial direction
(in contrast to a dynamical critical exponent of $1$,  that is  obtained for  many quantum critical spin models with only short-ranged interactions)~\cite{Dutta01}.
For $\sigma=1$, this gives a mean-field value of $z_\mathrm{MF}=1/2$.
With respect to the anomalous exponent $\eta$ that characterizes the spatial decay of the order parameter correlation function $G(r)$ at the quantum critical point,
different definitions are used in the literature. Here, we follow the standard notation, with  
\begin{equation}\label{eq:defeta}
G(r)\propto 1/r^{d+z-2+\eta}
\end{equation}
at criticality~\cite{Sachdev11}. According to Refs.~\onlinecite{Dutta01,Defenu17}, the value of $\eta$ in the relevant region of $\sigma$ for our study is fixed to $\eta=2-\sigma$, so that 
for $d=\sigma=1$, we obtain $G(r) \propto 1/r^{z}$. Again, for $d=\sigma=1$, this form agrees with the findings in Ref.~\onlinecite{Maghrebi16} (in their convention, $G(r)\propto 1/r^{d-1+\eta}$ and they obtain the relation $\eta=z$ for $\sigma=1$ and $n=1$). 
Another useful  result reported in Refs.~\onlinecite{Dutta01, Maghrebi16} for the case $\sigma=1$ of interest here,
is the relation 
\begin{equation}
\gamma/\nu=1
\end{equation}
between $\nu$ and the order parameter susceptibility  exponent $\gamma$, which we can access in our model by the single leg susceptibility $\chi_\mathrm{leg}$. 
Combined with the value of $\eta=2-\sigma$, this relation follows from the general scaling relation 
$\gamma=\nu(2-\eta)$.
In the following, we compare these RG results to our QMC estimates of the critical exponents, based on  a finite-size scaling analysis of the numerical data. 

For this purpose, we first shortly review the general  finite-size scaling theory near a quantum critical point. 
In particular,
for a  quantity $A$ that in the thermodynamic limit at $T=0$ scales as $A\propto \delta g^{\phi_A}$ with the relative deviation $\delta g = |g-g_c |/g_c$ from the quantum critical point at $g_c$, the corresponding finite-size scaling form in the  critical regime reads
\begin{equation}\label{eq:scalingansatz}
A(L,T,g) \propto L^{-\phi_A/\nu} \: F_A(\delta g\:  L^{1/\nu}, TL^z),
\end{equation}
in terms of  a scaling function $F_A$. 
In order to probe the critical properties near $g_c$ based on finite-temperature simulations, one  performs
low-temperature simulations for different system sizes $L$, scaling the inverse temperature $1/T \propto L^z$. One  can then perform the scaling analysis in terms of a single scaling variable, since  the second argument, $TL^z$, of $F_A$ then takes on a constant value.  
Based on the above estimate for the dynamical critical exponent $z$, we set the simulation temperature to $T=T_L$, where $T_L$ scales  as $1/T_L=2 L^{z}$  in order to reach the quantum critical scaling regime near $g_c$ (below we also determine  an estimate of $z$ that compares well to  the RG predictions).  
From Eq.~(\ref{eq:scalingansatz}), we see that the rescaled data sets of $A(L,T_L,g) L^{\phi_A/\nu}$  for different system sizes $L$,  when plotted as  functions of $g$, exhibit a crossing point at $g=g_c$. Furthermore, one obtains a data collapse upon plotting the rescaled values of $A(L,T_L,g)   L^{\phi_A/\nu}$ 
for different  system sizes as functions of $\delta g\: L^{1/\nu}$  for $g$ near $g_c$. These standard analysis techniques will now be used in order to estimate $g_c$ as well as the critical exponents for the Hamiltonian $\tilde{H}$ in the following. 

\begin{figure}[t]
\includegraphics[width=\columnwidth]{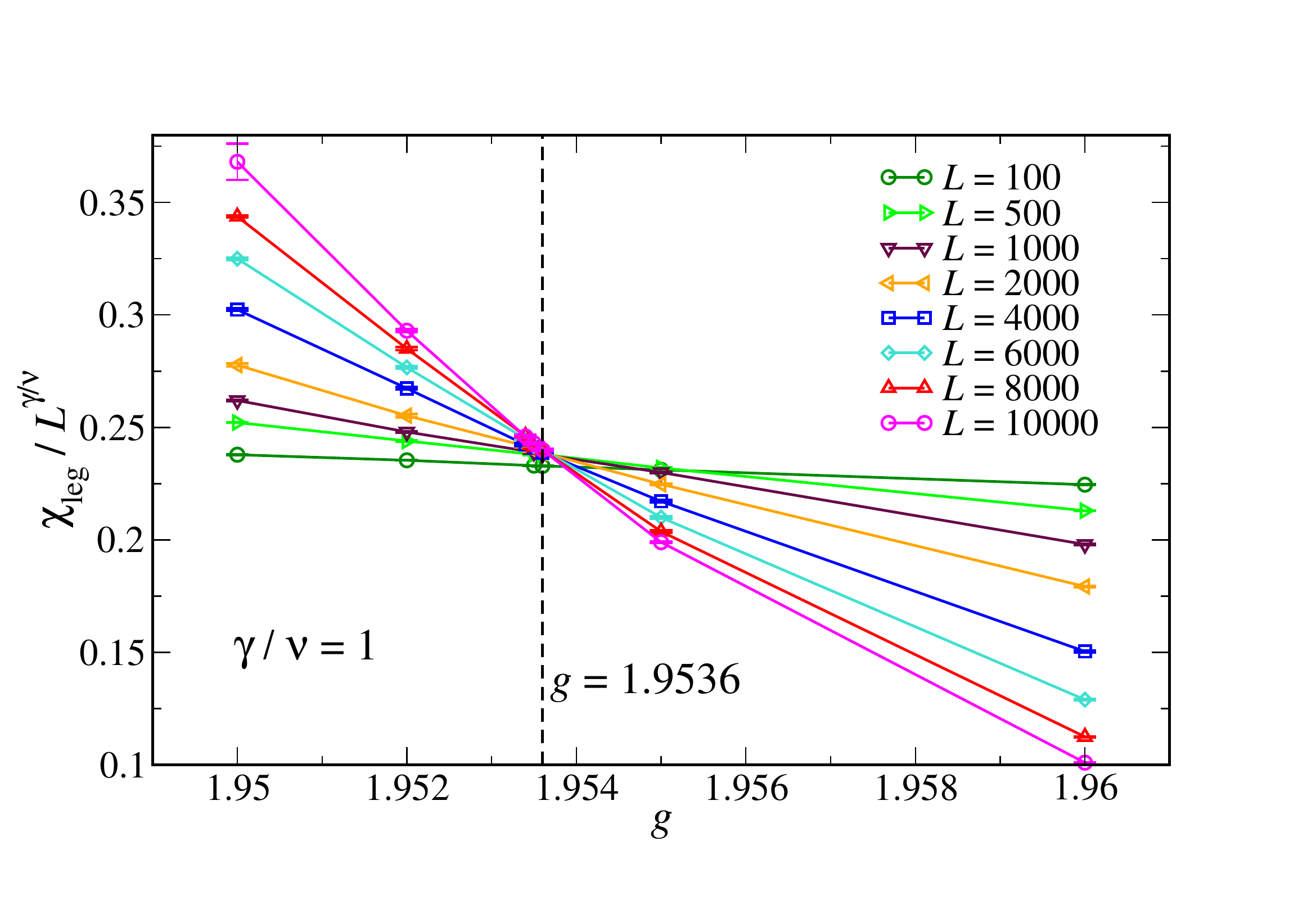}
\caption{(Color online) Crossing point analysis of the single-leg susceptibility $\chi_\mathrm{leg}$.}
\label{fig9}
\end{figure}

\begin{figure}[t]
\includegraphics[width=\columnwidth]{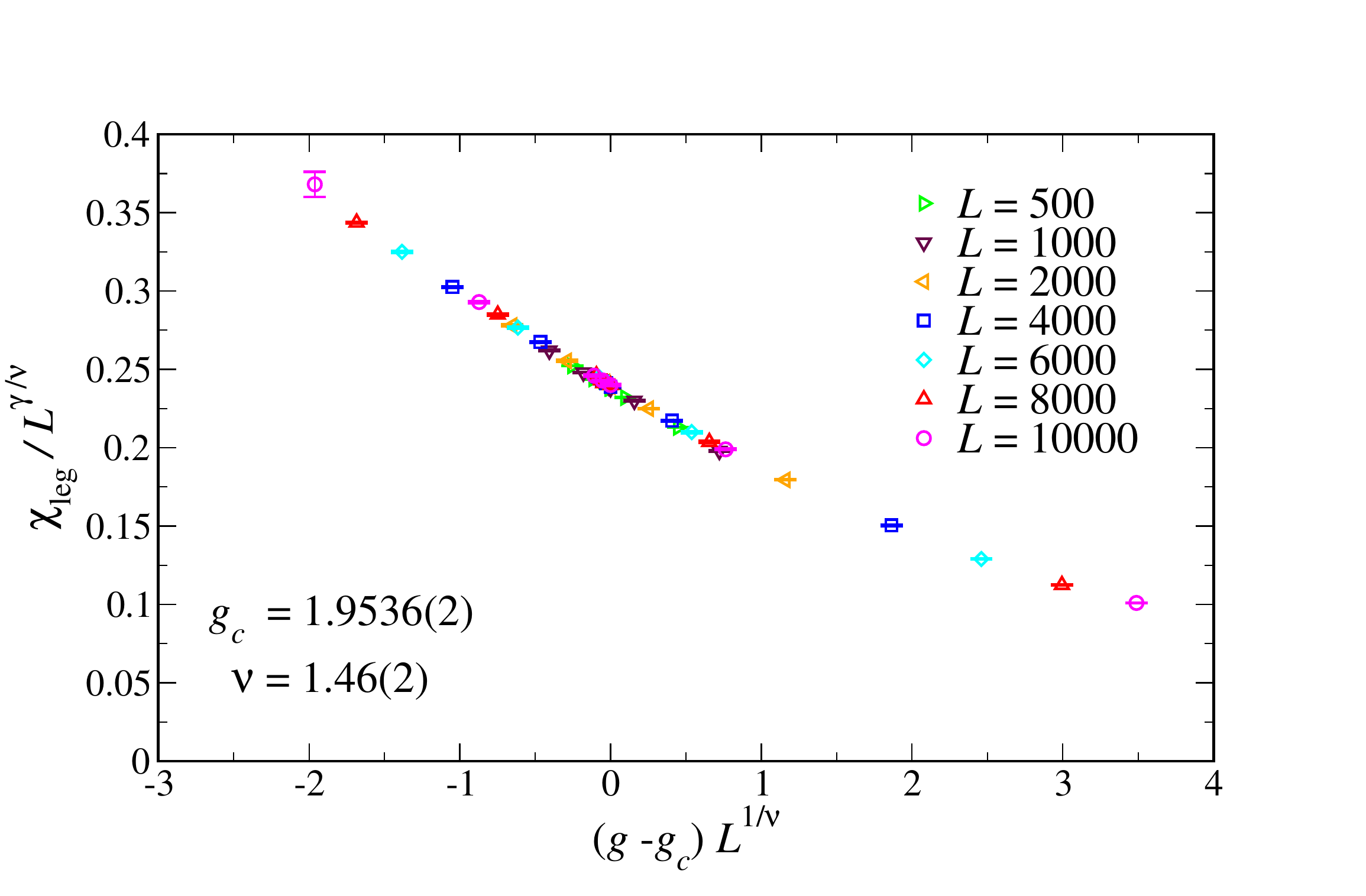}
\caption{(Color online) Data-collapse plot for the single-leg susceptibility $\chi_\mathrm{leg}$.}
\label{fig10}
\end{figure}

In our system, the order parameter quantifies the ferromagnetic alignment within each single leg, and the corresponding susceptibility is given in terms of  
the single-leg susceptibility $\chi_\mathrm{leg}$, which we  examined already  in the previous section. Here, we consider in more detail its finite size scaling. Using the RG prediction of $\gamma/\nu=1$, we indeed observe a crossing point in a plot of the rescaled finite-size data $\chi_\mathrm{leg}/L^{\gamma/\nu}$ as a function of $g$, cf. Fig.~\ref{fig9}.
We observe a  sharp crossing of the finite-size data for sufficiently large values of $L\geq500$.
Only the data for the smallest shown system size,  $L=100$,   exhibits the presence of further corrections to scaling. 
This crossing plot allows us to obtain a refined value of $g_c \approx 1.9536$. Furthermore, from a corresponding data collapse plot of the data for $L > 1000$, we obtain the estimates $g_c=1.9536(2)$ and $\nu=1.46(2)$, cf. Fig.~\ref{fig10}. 
In agreement with the above RG-based estimate $(\nu\approx1.227)$, our result for $\nu$ is larger than the mean-field value~\cite{Dutta01} $\nu_\mathrm{MF}=1$ for $\sigma=1$. Our numerical value for $\nu$ extends beyond the  RG-based estimate, which however was extrapolated from only the linear-order expression in $\epsilon$. 
It would of course be valuable to have at hand 
 more accurate RG analytical estimates for $\nu$ to compare with. 

\begin{figure}[t]
\includegraphics[width=\columnwidth]{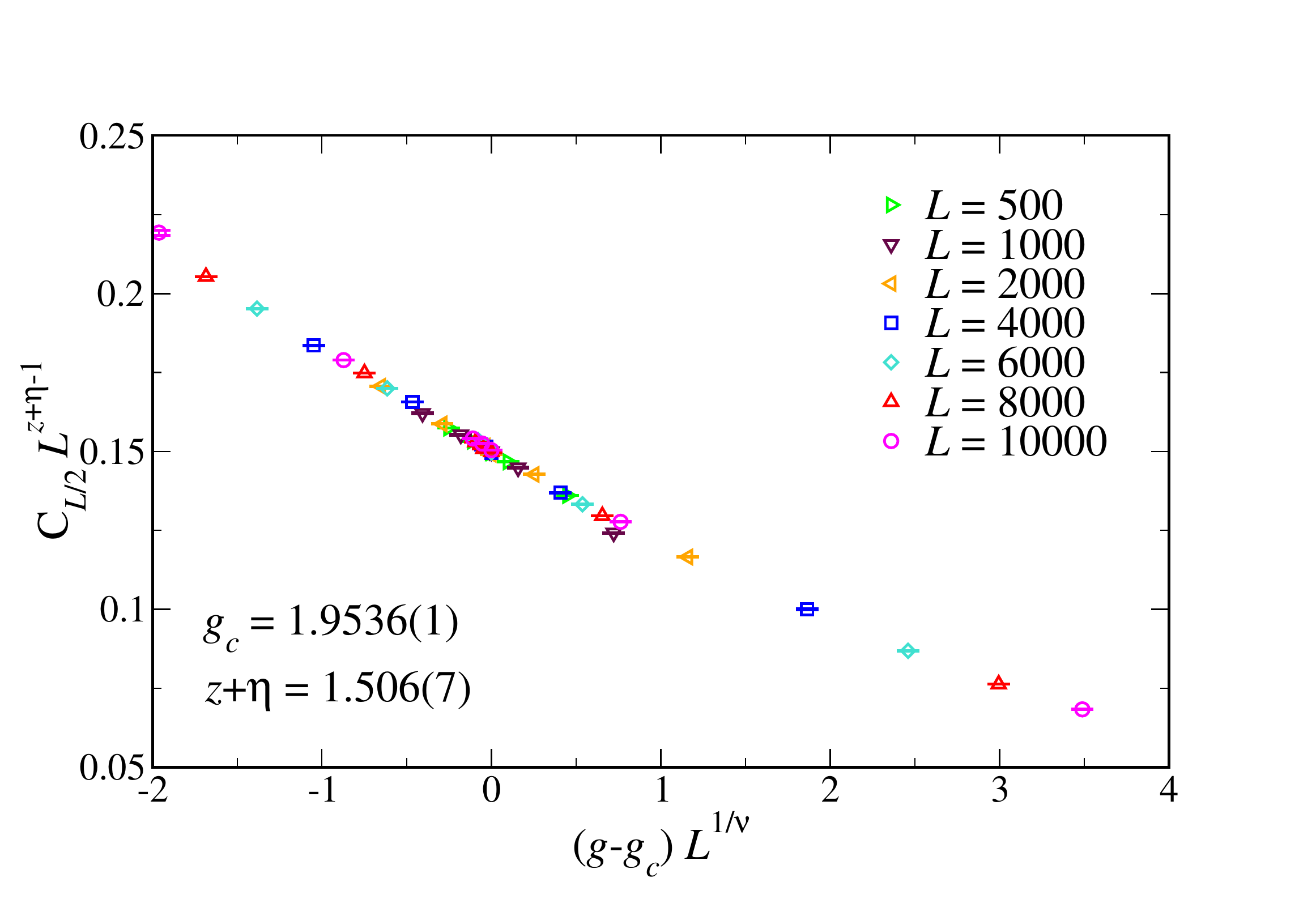}
\caption{(Color online) Data-collapse plot for the long-distance correlations $C_{L/2}$.}
\label{fig11}
\end{figure}

\begin{figure}[t]
\includegraphics[width=\columnwidth]{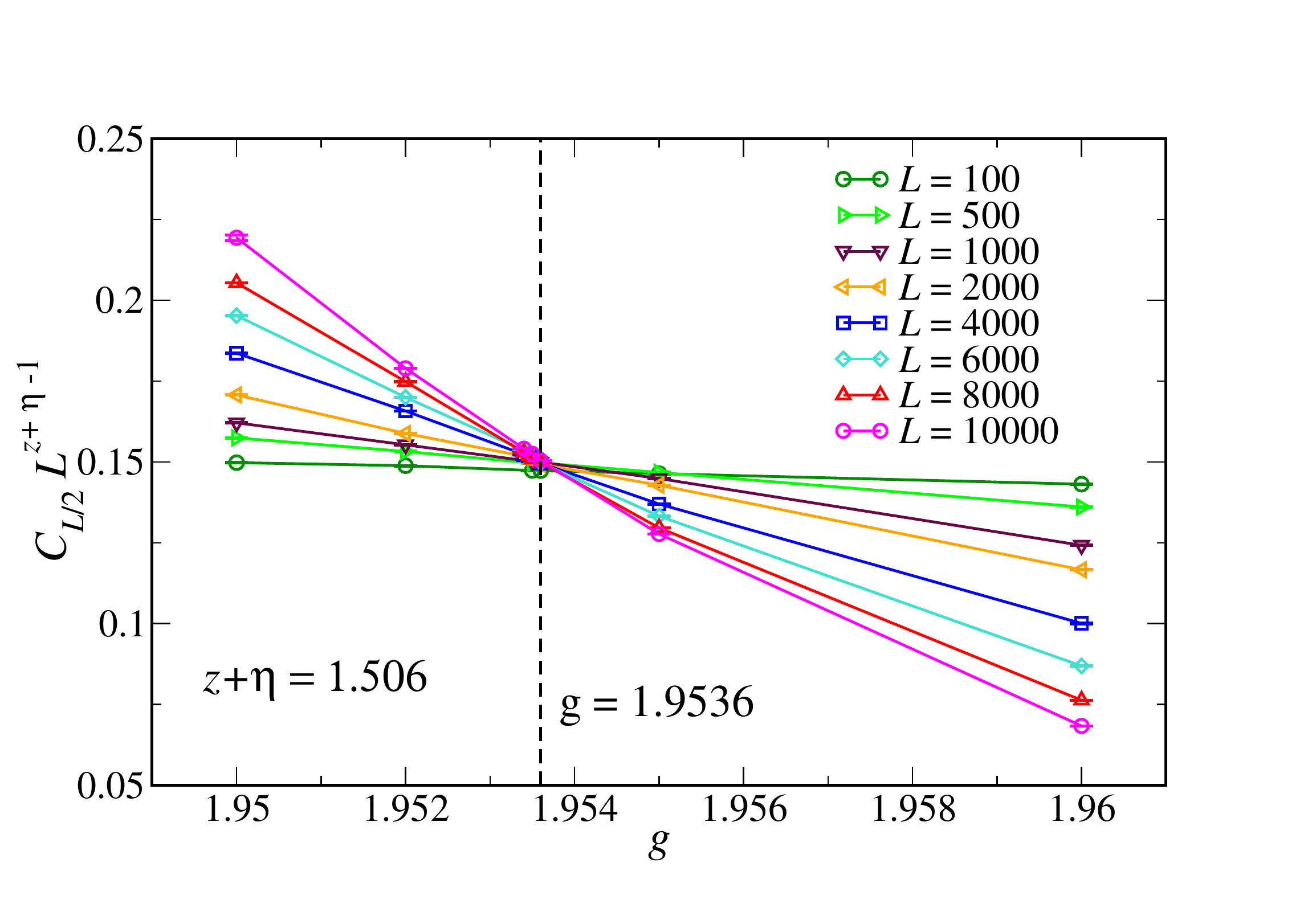}
\caption{(Color online) Crossing point analysis of the long-distance correlations $C_{L/2}$.}
\label{fig12}
\end{figure}

In order to  directly access the long-distance intra-leg correlations, we  measured in the QMC simulations the correlations between spins on the same leg at the largest accessible distances (under PBC) for a given system length $L$.
By averaging  over the values of the correlations at   $l$   distances  around the maximum distance $L/2$ for a given lattice size $L$, we  obtain a better statistics on this quantity, which we denote by $C_{L/2}$, and where we used a value of $l=0.01L$.  
Based on Eq.~(\ref{eq:defeta}) with $d=1$, 
at criticality $C_{L/2}$ scales as $1/L^{z+\eta-1}$ with the system size $L$. 
A corresponding data collapse plot, using our above estimate of $\nu$, is shown in Fig.~\ref{fig11}, and allows us to infer a value of $z+\eta=1.506(7)$, and a value of $g_c=1.9536(1)$, which matches well with the above estimate. Furthermore, we observe a corresponding crossing point in the rescaled data, cf. Fig.~\ref{fig12}.
When combined with the relation $\eta=2-\sigma$, we obtain from this analysis a value of $z=0.506(7)$. Note that this result is also in accord with the overall algebraic decay of $C(r)$ near the quantum critical point  observed in Fig.~\ref{fig4}.

\begin{figure}[t]
\includegraphics[width=\columnwidth]{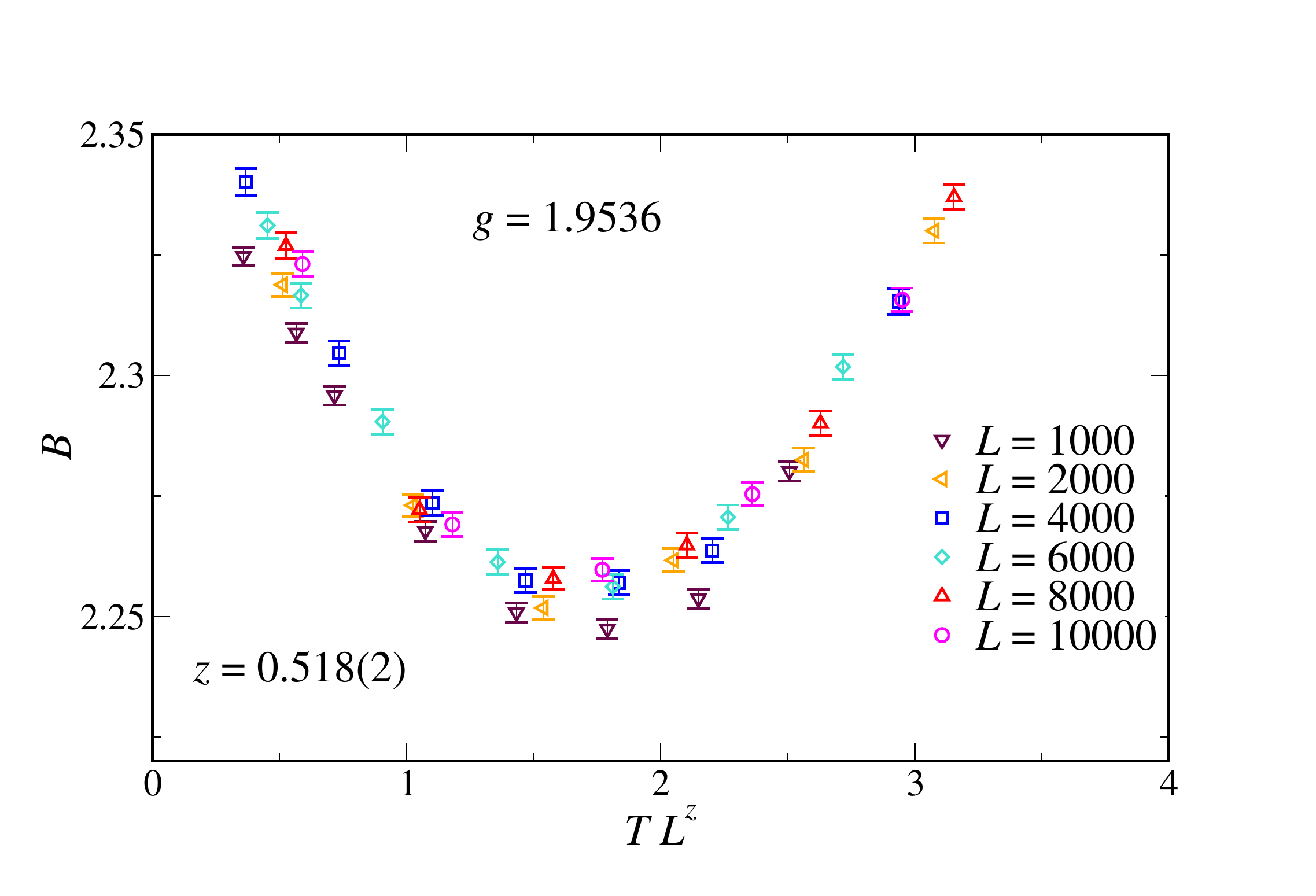}
\caption{(Color online) Data-collapse plot for the Binder ratio $B$ from simulations at finite temperatures atop the estimated quantum critical point.}
\label{fig13}
\end{figure}

We can furthermore obtain a separate estimate of the dynamical critical exponent $z$ by performing finite-temperature simulations
within the quantum critical region top $g=g_c$.
This is in particular convenient, since we actually use a finite-temperature QMC simulation method. 
In particular, we consider for this purpose the Binder ratio for the single-leg magnetic moment, 
 \begin{equation}
 B=\frac{\langle (M_\mathrm{leg})^4 \rangle }{\langle (M_\mathrm{leg})^2\rangle^2}.
 \end{equation}
For finite temperatures  within the quantum critical region atop the quantum critical point,  this dimensionless quantify ($\phi_B=0$) scales as 
\begin{equation}
B(L,T, g_c)=F_B(TL^z),
\end{equation}
so that  a corresponding data collapse plot allows us to estimate the value of $z$ given our above estimate for $g_c$. 
Such a  collapse plot for the Binder ratio is shown in Fig.~\ref{fig13}, and we obtain from this an estimate of $z=0.518(2)$, 
which agrees  with the above value, given  the statistical uncertainty.   

\begin{figure}[t]
\includegraphics[width=\columnwidth]{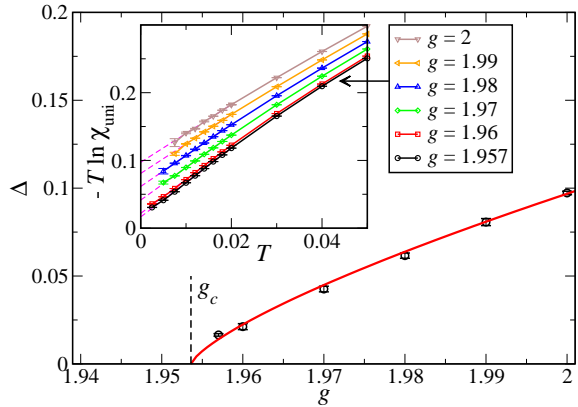}
\caption{(Color online) Softening of the spin excitation gap $\Delta$ in the quantum disordered region near the quantum critical point. Circles are  QMC estimates of the $\Delta$, and  the  solid line is a fit to the quantum critical scaling form, $\Delta \propto (g-g_c)^{z\nu}$, with $z\nu=0.739$, and $g_c=1.9536$ (indicated by the dashed line). The inset shows the low-temperature behavior of the uniform susceptibility $\chi_\mathrm{uni}$ for the $L=8000$ system along with  linear extrapolations 
(dashed lines) for temperatures $T<0.02$, in order to extract the gap $\Delta$ as the extrapolated value  of $ -T \ln\chi_\mathrm{uni}$ at $T=0$.  }
\label{fig14}
\end{figure}

Finally, we examine the scaling of the spin excitation gap $\Delta$ in the quantum-disordered phase close to the quantum critical point. We obtain an estimate for $\Delta$ from a fit of the low-temperature susceptibility $\chi_\mathrm{uni}$  to the  leading low-$T$ expression for  an activated behavior,
$
\chi_\mathrm{uni}\sim e^{-\Delta/T}
$.
We performed a linear regression of the corresponding  linear temperature dependence of 
$-T \ln \chi_\mathrm{uni}$, using the data for $\chi_\mathrm{uni}$ for $T<0.02$, in order  to estimate $\Delta$ for  values of $g$ close to $g_c$. This procedure is shown in the inset of Fig.~\ref{fig14}, based on the $L=8000$ data. 
Furthermore, 
near the quantum critical point, the spin gap is expected to scale as 
\begin{equation}
\Delta \propto (g-g_c)^{z\nu}.
\end{equation}
In the main panel of Fig.~\ref{fig14}, we show our results for $\Delta$ in the vicinity of the quantum critical point, along with a fit to this scaling form, based on a value of $z\nu \approx 0.739$, as  extracted from our above estimates for the two involved critical exponents.  The scaling form fits the numerically estimated $g$-dependence of the gap rather well. The weakly larger value of $\Delta$ extracted for the point closest to $g_c$, as compared to the scaling form,  indicates  finite-size corrections near criticality. These are however anticipated, given that our QMC estimates for $\Delta$ are based on finite-system ($L=8000$) data. 
Overall, our numerical analysis  thus confirms the 
presence of a quantum critical point with an emerging scaling behavior for the effective spin model $\tilde{H}$, separating a gapless low-$g$ phase from the gapped large-$g$ quantum disordered regime. 
 
\section{Discussion}\label{sec:discussion}

In the preceding section we  observed, based on quantum Monte Carlo simulations combined with a finite-size scaling analysis, that both effective spin-ladder models which we 
considered here exhibit a quantum phase transition between a gapless, weak inter-leg coupling regime with a finite ferromagnetic  polarization  within each leg, and a gapped, strong inter-leg coupling quantum disordered phase.
While the regime of strong inter-leg coupling  is dominated by the formation of rung-based singlets, similar to the two-leg ladder with short-ranged interactions~\cite{Kolezhuk96,Roji96,Vekua03}, the
weak-coupling phase is in fact more appropriately understood in terms of two antiferromagnetically coupled superspins, each forming along one of the legs. In this sense, the basic picture of edge magnetism in graphene nanoribbons  is  apparently appropriate
for the effective quantum spin model in the relevant parameter region for sufficiently wide zigzag nanoribbons. 
Note that
due to the bipartite nature of the coupling geometry,  Marshall's theorem  implies that for any  finite ladder (i.e., finite $L$),  the ground state  is a global spin-singlet ($S=0$)~\cite{Marshall55,Lieb62}.
This corresponds to the fact that 
for any finite (zigzag) nanoribbon the ground state  within the Hubbard model description is a singlet due to 
Lieb's theorem~\cite{Lieb89}. 

There are  thus two distinct phases in the thermodynamic limit, which are both in accord with the singlet nature of the finite-system ground state.
The situation in  the effective spin-ladder model is in fact closely related to more familiar cases such as, e.g., the Heisenberg model on the square lattice bilayer~\cite{Singh88,Millis93,Millis94,Sandvik94,Chubukov95,Sommer01,Wang06}: there, the system realizes a quantum disordered phase for strong inter-layer coupling , and an antiferromagnetic phase with finite sublattice polarizations for weak inter-layer coupling.  However, and in contrast to the bilayer case, (i)  the two polarized sublattices of the effective ladder systems considered here are well separated from each other in real space,  and (ii)  direct, long-ranged ferromagnetic intra-leg couplings are required in order to stabilize the weak-coupling phase, given the reduced dimensionality of the effective spin-ladder systems.

\begin{figure}[t]
\includegraphics[width=\columnwidth]{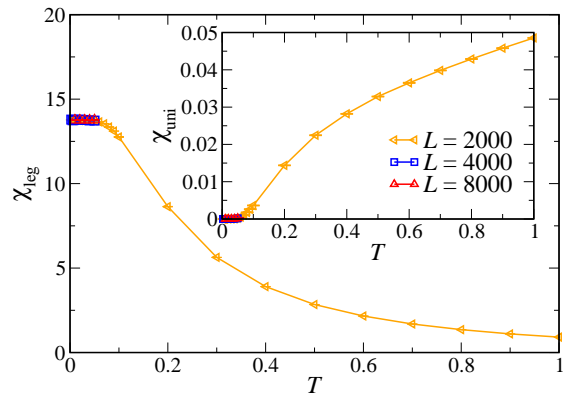}
\caption{(Color online)  (Color online) Temperature dependence of the single-leg susceptibility $\chi_\mathrm{leg}$ 
(main panel) and the  uniform susceptibility $\chi_\mathrm{uni}$
(inset) of the effective quantum spin model for a $W=6$ zigzag nanoribbon at $U=t$ for different values of $L$. }
\label{fig15}
\end{figure}

We found that the coupling parameters of the effective spin-ladder model derived
from the Hubbard model description for the width $W=10$ zigzag nanoribbon~\cite{Schmidt13, Koop15} locate the corresponding effective quantum spin model well within the weak-coupling region.  The quantum disordered region  is reached only upon artificially enhancing the inter-leg coupling beyond the quantum critical coupling strength. 
For  even wider nanoribbons, the antiferromagnetic inter-leg couplings of the effective ladder model will be further reduced~\cite{Schmidt13, Koop15}, so that also for such nanoribbons the effective spin model   resides within the weak-coupling regime. This is in  contrast to the previously considered case of chiral nanoribbons, for which the effective spin models  
had a quantum disordered, spin-gapped ground state~\cite{Golor14}. 
One may ask, whether instead the ground states of the spin-ladder models for narrower zigzag nanoribbons, with $W<10$, for which the antiferromagnetic inter-leg coupling is indeed larger,  reside within the gapped,  quantum disordered regime. 
In fact, previous numerical  studies of the extremely narrow $W=2$  zigzag nanoribbon, performed directly within the Hubbard model description, clearly identified a gapped
quantum disordered ground state~\cite{Hikihara03,Feldner11}. 
Motivated by the observation of a quantum phase transition in the effective spin models for the $W=10$ nanoribbon, we  performed quantum Monte Carlo simulations also for the effective spin-ladder model for a $W=6$ zigzag nanoribbon (again for $U=t$), even though such a ribbon may   already be too narrow for the effective spin model derivation to still be applicable~\cite{Koop15}.
For the resulting effective spin-ladder model for the $W=6$ nanoribbon, we  obtain a ratio of $J^{AF}(r=0)/J_F= 6.141$ between the nearest-neighbor antiferromagnetic inter-leg coupling and the long-ranged ferromagnetic intra-leg tail,  which is significantly larger than the corresponding ratio of 
$1.95$ for the $W=10$ nanoribbon. 
Figure ~\ref{fig15} shows the temperature dependence of both the uniform susceptibility and the single-leg susceptibility of the effective spin-ladder model for the $W=6$ nanoribbon as obtained from the QMC simulations,  exhibiting that this effective spin-ladder model indeed has a gapped, quantum disordered ground state.  From the temperature dependence of the uniform susceptibility, we estimate a corresponding spin gap of $\Delta\approx 0.3 \: J^\mathrm{F}(1)= 0.014t$, in terms of the Hubbard model hopping strength (we also considered explicitly the case of $W=8$, for which we find the  ground state to be located within the gapless, weak-coupling region). 
Even though the truncated effective spin model derivation will be less accurate for such a narrow ribbon, the above result shows that indeed  both phases may in principle be  accessed in  effective spin-ladder models for zigzag graphene nanoribbons.

Anticipating the fact that  the
effective quantum spin models  describe the correlations among the edge magnetic moments in graphene nanoribbons
within a controlled, but nevertheless approximate framework, we  are  not in a position  to discern, based on our findings,
whether  edge magnetism is indeed stabilized in wide graphene  zigzag nanoribbons, at least in the ground state. 
For this purpose, 
various additional effects may also have to be considered,  such as electronic interactions beyond the local Hubbard repulsion~\cite{Shi17},
extended hopping terms in the kinetic energy, as well as exchange anisotropies deriving, e.g.,  from  graphene-to-substrate couplings~\cite{Zhang16}.   
The above finding nevertheless
represent a plausible scenario for stable edge-magnetism on wider zigzag nanoribbons, at least within the effective spin model for the the most-basic Hubbard model description. 
It  would  be worthwhile to extend beyond our investigation towards analyzing also the  low-energy spin dynamics  and  its evolution across the quantum phase transition, which is feasible, e.g., with quantum Monte Carlo methods. 
Moreover, 
the real-time out-of-equilibrium behavior of such effective spin models with long-ranged interactions can be probed in order to examine the quantum nature of the spin response, and the evolution of the relevant time-scales of the magnetic fluctuations~\cite{Golor14} both within  the weak-coupling regime as well as upon crossing the quantum critical point. Such a study could be performed using advanced numerical methods for quantum  systems with long-ranged interactions~\cite{Zaletel15}, and is also left for future investigations.

\section*{Acknowledgments}
We thank  M. Golor, F. Hajiheidari, R. Mazzarello, and M. J. Schmidt for useful 
discussions
and acknowledge support by the Deutsche Forschungsgemeinschaft (DFG) under grant FOR 1807 and RTG 1995. Furthermore, we thank the IT Center at RWTH Aachen University and the JSC J\"ulich for access to computing time through JARA-HPC.
%

%
%
\end{document}